\documentstyle[preprint,aps]{revtex}
\begin{document}
\preprint{IP/BBSR/95-12; hep-th/9504020}
\title { On chiral symmetry breaking by external magnetic
fields in QED$_{3}$}
\author{ Rajesh R. Parwani}
\address{Institute of Physics, Bhubaneswar-751 005,
INDIA.\footnote{e-mail: parwani@iopb.ernet.in }}
\maketitle
\begin{center}
Feb 1995 (Revised May 1995)
\end{center}
\begin{abstract}
A recent result of Gusynin, Miransky and Shovkovy concerning
chiral symmetry breaking by a constant external magnetic field
in parity-invariant three-dimensional QED is generalised to the
case of inhomogeneous fields by relating the phenomenon to the
zero modes of the Dirac equation. Virtual
photon radiative
corrections and four-dimensional
QED are briefly discussed.

\end{abstract}
\vspace{2 cm}
%\pacs{PACS NO.12.20.Ds, 11.30Qc, 03.65Ge. }
\narrowtext

\newpage

%\section {Introduction}
%\label{int}
In a recent paper Gusynin, Miransky and Shovkovy \cite{GMS}
showed that a constant
external magnetic field causes chiral symmetry breaking in three
dimensional parity-invariant quantum
electrodynamics (QED$_3$). They further
demonstrated the same effect in the Nambu-Jona-Lasinio (NJL)
model to support their contention that the phenomenon was
universal in $2+1$ dimensions (see also \cite{KK} for an earlier
study in the NJL model).

As lower dimensional field theories serve as simplified
models in particle physics, and also as effective theories
in condensed matter physics \cite{Frad}, the
phenomenon studied in Ref.\cite{GMS} deserves closer
examination.
In this Letter, the underlying ingredients
responsible for the field induced symmetry breaking in the QED
system of \cite{GMS} are exposed, and the result
generalised,  qualitatively and
quantitatively, to the case of inhomogeneous {\it external} magnetic
fields. The generalisation is desirable as it enhances the
potential phenomenological applicability \cite{GMS} of the results.
Near the end of this paper the other topics mentioned in the abstract
will also be discussed, and a remark made on the
NJL model.

The theory under consideration is defined here by the
Lagrangian density
\begin{equation}
{\cal{L}} = \bar{\Psi} \left( i D_\mu \ \Gamma^{\mu} - m \right)
\Psi \; ,
\label{lag}
\end{equation}
where $D_\mu = \partial_\mu + ie A_\mu$ and the {\it four}-component
spinor $\Psi$ forms a reducible representation of the Dirac
algebra
\begin{equation}
\Gamma^{0}=\pmatrix{\gamma^{0} & 0 \cr  0 & -\gamma^0} \ , \;\;\;
\Gamma^{1}=\pmatrix{\gamma^{1} & 0 \cr  0 & -\gamma^1} \ , \;\;\;
\Gamma^{2}=\pmatrix{\gamma^{2} & 0 \cr  0 & -\gamma^2} \ ,
\end{equation}
with $\gamma^0 = \sigma^3, \ \gamma^1 =i\sigma^1, \ \gamma^2
=i\sigma^2$. Defining \cite{PAK}
$ \Psi \equiv \pmatrix{\Psi_{+} \cr \Psi_{-}}$, with $\Psi_\alpha
\;\ (\alpha = \pm )$
two-component spinors, the Lagrangian density (\ref{lag})
is invariant under
the generalised
parity operation $(x,y) \to (-x,y), \ \Psi_{+} \to \sigma^1
\Psi_{-}$ and  $\Psi_{-} \to \sigma^1 \Psi_{+}$ .
Furthermore,
when $m \equiv 0$, it
is also
invariant under the chiral $U(2)$ symmetry generated by $I, \Gamma^5,$
and $-i\Gamma^3$, where $\Gamma^3 \equiv \pmatrix{0&i \cr i &0}$
and $\Gamma^5 =i \Gamma^0 \Gamma^1 \Gamma^2 \Gamma^3$.
In terms of two-component spinors,
Eq.(1) may be written as
\begin{eqnarray}
{\cal{L}} &=& \sum_{\alpha =\pm} {\cal{L}_{\alpha}} \equiv
\sum_{\alpha =\pm}
{\bar{\Psi}}_{\alpha} \left( i D_\mu \ \gamma^{\mu} -
\alpha \ m \right)
\Psi_{\alpha} \ .
\label{lag2}
\end{eqnarray}

One sees that in the one-fermion loop approximation (of which
the external
field problem is a special case),
(\ref{lag2}) describes two decoupled systems described by
$\cal{L}_{\pm}$. Taken separately, $\cal{L}_{\pm}$
describe the usual two inequivalent irreducible
representations of parity-noninvariant \cite{parity}
QED$_{3}$.
Restricting our attention to the
external field case from now on, the
fermion condensate is
\begin{eqnarray}
S(x;m) \equiv \langle\bar{\Psi}(x) \Psi(x) \rangle
&=& \sum_{\alpha = \pm} \alpha \ {\langle{\bar{\Psi}}_{\alpha}(x)
 \Psi_{\alpha}(x) \rangle}_{\alpha} \label{ord} \\
&\equiv& \sum_{\alpha= \pm } \alpha \ S_\alpha (x;m) \; ,
\label{cond}
\end{eqnarray}
where the notation ${\langle{\bar{\Psi}}_{+}
\Psi_{+} \rangle}_{+}$ means the expectation value of
${\bar{\Psi}}_+ \Psi_+$ in the $\cal{L_+}$ subsystem.
Since it follows from (\ref{lag2}) that $S_{-}(x;m) =
S_{+}(x;-m)\ , $
so (\ref{cond}) becomes
\begin{eqnarray}
S(x;m) &=& \sum_{\alpha = \pm} \alpha S_{+}(x; \alpha \ m)
\label{cond2} \, .
\end{eqnarray}
Thus in the external field approximation, the properties of the
condensate $S$ are determined completely by the condensate $S_+$
in the subsystem $\cal{L}_+$. Equation (\ref{cond2}) implies
\begin{eqnarray}
S(x;-m) &=& -S(x;m) \label{sym} \, ,
\end{eqnarray}
and therefore if $S$ were a continuous function of $m$ at $m=0$
then $S(x;m=0)=0$. In actuality, as will be soon be apparent,
$S(x;m)$  is discontinuous at
$m=0$ so that different, and in general  nonvanishing, values
for it are obtained in the opposing limits $m\to 0^{\pm}$.
Note that from the definition (\ref{ord}), a nonzero value
for $S(x;m \to 0)$ indicates chiral-symmetry breaking.

Physically, the quantity ${1 \over 2} S_{+}(x;m)= \langle
\Psi^{\dag}_{+} { \sigma^{3} \over 2} \Psi_{+} \rangle$
is the induced spin-density corresponding to the
theory described by the Lagrangian $\cal{L}_{+}$. For a
{\it static magnetic field}
 with corresponding flux $F= {e \over 2 \pi}
\int d^2 x B $, the net {\it induced} spin due to vacuum
polarisation, in the massless
limit, is a topological invariant given by \cite{BBY,Pol}
\begin{eqnarray}
\int d^2 x {S_{+}(x;m\to0) \over 2} &=& -{1 \over 4}
|F| \ \mbox{sign}(m) \; . \label{index}
\end{eqnarray}
Some technical points relating to (\ref{index}), and
which are of some importance, should be
noted. Firstly, since spin is charge-conjugation even
($C$-even), the {\it induced} vacuum spin on the right-hand-side
of Eq.(\ref{index}), has been obtained by taking
the $C$-even part of the quantity
on the left-hand-side after subtraction of an infinite
bare (zero field) vacuum contribution. Secondly,
if one had started with $m \equiv 0$ from the beginning then,
since massless electrons in $2+1$ dimensions are spinless, one
would have expected $S_{+}(x;m \equiv 0) =0$ thereby leading
through Eq.(\ref{cond}) to $S(x;m\equiv 0)=0$ in agreement
with (\ref{sym}).
However if a fermion
mass is introduced as an infrared regulator (as is the case
here----more on this at the end), then removing the
mass at the end of the calculation
gives a nonzero result;  from the
point of view of symmetries, this
resulting nonzero value for $S_+$ in
the massless limit is due to the fact that an explicit mass term
in $\cal{L}_{+}$ breaks parity which, being a
discrete symmetry, is not recovered in the continuous limit $m
\to 0$ (see, for example,\cite{BBB} and references therein).
Secondly, the appearance of
$\mbox{sign}(m)$ in Eq.(\ref{index})
is due to the above-mentioned fact that
the $m >0$ and $m<0$ cases correspond to inequivalent
representations (and thus different physical situations).

Let me now sketch a derivation of (\ref{index}) which
will reveal some information useful for later use.
Construct the eigenstates $\psi_E$ of the Hamiltonian $H_+$,
corresponding to the Lagrangian $\int d^2x \cal{L}_{+}$,
 for static
magnetic fields in the $A_0 =0$ gauge, and with $F > 0$. Then
in the $m \to 0$ limit  the $E >0$ and $E<0$ eigenstates are
related \cite{ACJ}
by $\psi_{-E} = \sigma^{3} \psi_{E}$, so that their
net contribution to
\begin{eqnarray}
S_{+}(x;m\to 0) &=& {\lim_{m \to 0}}
\langle {\bar{\Psi}_{+}} \Psi_{+} \rangle |^{F}_{0} \nonumber \\
 &= & -{1 \over 2} {\lim_{m \to 0}} \sum_{E} \mbox{sign}(E)
\psi_{E}^{\dag} \sigma^{3} \psi_{E} |^{F}_{0}
\end{eqnarray}
vanishes. The only unpaired states are the
zero modes at $E=m \to 0$ for the $F> 0$ case and at
$E=-m \to 0 $ for the $F < 0 $ case. These zero modes are
of two types: Denoting by $[F]$ the smallest integer greater
than or equal to $|F|-1$, there are $[F]$ normalisable
(to unity) states
in addition to the resonant (scattering) states \cite{ACJ,Pol}.
Since the
zero modes are of the form $\psi_{E=m \to 0} \sim \pmatrix{u \cr 0}$
and $\psi_{E=-m \to 0} \sim \pmatrix{0 \cr v}$
it follows that
\begin{eqnarray}
{1 \over 2} \ S_{+}(x;m\to 0^{\pm}) &=& -{ \mbox{sign}(m) \over
4} \sum_{E=0}
\psi_{E}^{\dag}  \psi_{E} |^{F}_{0} \;\, , \;\;\;\;\;\;
\, . \label{zer}
\end{eqnarray}
Integrating the right-hand-side of (\ref{zer}) over all space
shows that the normalisable states contribute an amount
$-{\mbox{sign}(m) \over 4}  [F] $, while a detailed analysis
\cite{Pol}
shows that the
resonant states contribute $-{\mbox{sign}(m) \over 4} (|F|-[F])$.
Thus   one has
\begin{eqnarray}
\int d^2 x { S_{+}(x;m \to 0^{\pm}) \over 2} &=&
-{\mbox{sign}(m) \over 4} \ |F| \,
\end{eqnarray}
which is  the result quoted in Eq.(\ref{index}).

The Eqs.(\ref{cond2}), (\ref{index}) and (\ref{zer})
are the main points
in this paper from which several deductions may be made
about the chiral-symmetry breaking order parameter given by
$\langle {\bar{\Psi}} \Psi \rangle_{m \to 0}$.
Combining (\ref{cond2}) and (\ref{index}) one obtains the first
result
\begin{eqnarray}
\int d^2x {\langle\bar{\Psi} \Psi \rangle}_{m \to 0}
 &=& -|F| \ \mbox{sign}(m) \, , \label{res1}
\end{eqnarray}
showing that the spatial average of the order parameter is a
topological invariant depending only on the net flux, and is
nonvanishing if the net flux is nonzero.
For a uniform field $B$, translational invariance applied to
(\ref{res1}) then implies
 \begin{eqnarray}
 {\langle\bar{\Psi} \Psi \rangle}_{m \to 0}
 &=& -{|eB| \over 2\pi } \mbox{sign}(m) \, , \label{const}
\end{eqnarray}
which agrees with the result of Ref.\cite{GMS} obtained
by an explicit
calculation using the Schwinger proper-time method (in
\cite{GMS} it was implicitly assumed  that $m>0$ so that no factor
of $\mbox{sign}(m)$ is visible there).

Next, if the magnetic field is very slowly varying, one guesses
from (\ref{res1}) and dimensional analysis that correction terms
of the form  $ \partial (eB)^{ 1 \over 2}$ and
$\partial^2(eB) \over eB$ might occur on the right-hand-side
of (\ref{const}). This suggestion could perhaps be checked by an
explicit calculation along the lines of a recent gradient expansion
of the effective action in Ref.\cite{CDD}. Much more definitive
statements about the order parameter, which apply even to
field configurations deviating substantially from
homogeniety (such as localised vortices), can be made by using
the unaveraged version of (\ref{index}) given by (\ref{cond2}) and
(\ref{zer}). From the discussion leading up to (\ref{zer})
it is deduced that
${\langle{\bar{\Psi}} \Psi \rangle}_{m \to 0}$ is
concentrated near the vortex itself since \cite{ACJ} that is
where the
normalisable zero modes are localised
(the resonant zero modes give an infinitesimal local
contribution \cite{Pol}).

A simple explicit example which illustrates some features
of inhomogeneous configurations is
provided by a thin flux ring : $B = {F \over r} \ \delta(r-R)$
with $F=N+\epsilon, \ N \ge 1$ and $0< \epsilon < 1$. Then in the
Lagrangian $\cal{L}_{+}$ with $m>0$ the $N$ normalisable states
at $E=m$ are given in polar coordinates by
\begin{eqnarray}
\psi_n &=& \pmatrix{u_n(r) \cr 0}\, , \;\; \;\;\; 1 \le n \le N \, ,
\nonumber \\
&& \nonumber \\
u_n(r) &=& { e^{i(n-1)\theta} \over \sqrt{\pi}} \ \left({n(F-n)
\over F} \right)^{1\over 2} \ g(r) \, ,
\end{eqnarray}
where $g(r) = r^{(n-1)}/R^n$ for $r \le R$ and $g(r) = R^{(F-n)}/
r^{(F-n+1)} $ for $r \ge R$.
These eigenstates decay  algebraically away from the localised
field, a feature
which is typical \cite{ACJ,Pol} of normalisable states around
a vortex.
As  these states are independent of the magnitude of $m$ they
remain localised around the vortex in the massless limit while,
as mentioned above, the resonant zero modes have their
contribution smeared over all space. When the radius $R$
is decreased,
the normalisable zero modes begin to stick to the ring and in
the extreme limit
$R \to 0$ these states collapse into
point-like states sitting
exactly on the infinitely thin flux string \cite{PG}.
Thus for an Aharonov-Bohm flux string with $F >1$ the
non-negligible contribution to the order parameter
${\langle {\bar{\Psi}\Psi} \rangle}_{m \to 0}$
is a delta-function support at the
string.
An explicit verification of
Eq.(\ref{res1}) for the $F < 1$ magnetic string
(when only the dilute resonant contribution is present) is
technically interesting,
 and will be presented elsewhere \cite{RP}.

To summarise the main points, chiral symmetry breaking by
external magnetic fields in the reducible but parity-invariant
representation of
$QED_3$ defined by $\cal{L}$ is related to the induced vacuum
spin-density in the underlying irreducible but
parity-noninvariant representation of $QED_3$ defined
by ${\cal{L}}_+$.
The induced spin-density in $\cal{L}_+$ (in the required
massless limit),  is
determined solely by the zero modes of the Dirac Hamiltonian.
Since the arguments presented here did not assume a constant
magnetic
field, the discussion of Ref.\cite{GMS} has been generalised with
Eqs.(\ref{cond2}), (\ref{zer}) and (\ref{res1}) providing the
main information. In particular, for a finite flux $|F|>1$,
the nonnegligible contribution to the local order parameter
comes only from the normalisable zero modes. If the flux
$|F|<1$, then locally the order parameter $\langle {\bar{\Psi}} \Psi
\rangle_{m \to 0}$ is essentially zero. The resonant zero modes
appear to be important locally only if the flux is infinite
(globally they of course contribute to the index (\ref{res1})).

Please note that the order parameter
depends on $\mbox{sign}(m)$ (as is typical of induced quantum
numbers in $QED_3$ \cite{NSR,BBY,MP,Pol});
thus  the situation may be described
as one in which the "direction" of field-induced dynamical
chiral symmetry breaking is determined by the "direction"
of an {\it infinitesimal} explicit
breaking ($m=0^{\pm}$). However if it were possible to
regulate the $m=0$ theory
(\ref{lag}) unambiguosly without explicitly
breaking the chiral symmetry,
then it would seem from (\ref{cond2}) that there should be no
chiral symmetry breaking induced by external fields.
In that (assumed) case, the
results obtained in \cite{GMS} and here would have to be
interpreted as saying that an {\it infinitesimal} explicit breaking
of the chiral symmetry is required
to seed a {\it finite} breaking of that symmetry in an
external magnetic field. On the
other hand, it is believed that in the $m=0$ theory (\ref{lag})
a dynamical fermion mass is generated
{\it nonperturbatively} \cite{PAK,ARY} even in the absence of
external
fields. In this latter scenario the results of \cite{GMS} and
this paper then suggest that an external magnetic
field amplifies the truly dynamical chiral symmetry breaking.

One can ask how radiative corrections affect
the results of Eqs.(\ref{cond2}), (\ref{zer}) and (\ref{res1}).
If the external field is strong ($eA_\mu \gg e^2$) then virtual
photon
corrections will be perturbatively suppresed. Even if the
external field is only moderately strong, the external field
topological index (\ref{res1}) makes it plausible that,
by approximately treating the virtual photons as
classical fluctuations of the external field, the fermion
condensate will
not be totally washed away. If the external field is very weak,
or absent, then radiative corrections can be important and a
self-consistent approximation scheme must be used \cite{PAK}.

So far the discussion above has been for $QED_3$.
On setting $m=0$ in $\cal{L}$ but adding a term
$G({\bar{\Psi}}\Psi)^2$
one obtains the NJL model which was  studied in depth
in Ref.\cite{KK} and also  examined  in
\cite{GMS}. Unfortunately for this case the
decompositions (\ref{lag2}) and (\ref{cond2})
no longer hold. It would clearly be useful if
in the NJL case too a
picture for the chiral symmetry breaking by magnetic fields can
be achieved which allows a generalisation to inhomogeneous
fields the results of \cite{KK,GMS}. I do not know of one.

Finally let me mention some
possible new phases of four-dimensional QED, suggested by studies
using external magnetic fields as probes. By examining
vacuum polarisation effects around thin flux tubes, it was
proposed in Ref.\cite{GLP} that there might exist a
nonperturbative strong-coupling phase of $QED_{4}$ with a vacuum
consisting of dynamical flux strings. On the other hand in
\cite{GMS2} it was argued that in the weak coupling phase of
$ QED_{4} $, a constant magnetic field induces chiral-symmetry
breaking. Whether these two pictures, among others in the
literature \cite{phases},
are compatible or
complementary is as yet unclear.

The problems of the last two paragraphs, the
extensions of the $QED_3$ analysis to nonzero temperature,
chemical potential and external electric fields, and possible
concrete
applications are open questions left for future investigations.

\acknowledgements
I thank Pijush Ghosh, Alfred Goldhaber, Avinash Khare and
Hsiang-nan Li for helpful conversations, and for a reading of the
manuscript.

\end{document}